\documentclass[%
 reprint,
 amsmath,amssymb,
]{revtex4-2}

\usepackage{amssymb}
\usepackage{amsmath}
\usepackage{amsthm}
\usepackage{graphicx}
\usepackage{dcolumn}
\usepackage{bm}
\usepackage{hyperref}
\usepackage[usenames,dvipsnames,x11names,table]{xcolor}
\usepackage{balance}
\usepackage{tikz}
\usetikzlibrary{shapes}
\usepackage{enumitem}
\setlist[itemize]{align=parleft,left=0pt..1em}

\begin{document}


\title{Liquid film rupture beyond the thin-film equation: a multi-component lattice Boltzmann study}

\author{Francesca Pelusi$^{1}$}
\email{f.pelusi@fz-juelich.de}
\author{Marcello Sega$^1$}
\author{Jens Harting$^{1,2}$}
\affiliation{$^1$Helmholtz Institute Erlangen-Nürnberg for Renewable Energy (IEK-11), Forschungszentrum Jülich GmbH, Cauerstraße 1, 91058 Erlangen, Germany}
\affiliation{$^2$Department of Chemical and Biological Engineering and Department of Physics, Friedrich-Alexander-Universität Erlangen-Nürnberg, Cauerstraße 1, 91058 Erlangen, Germany}

\vspace{0.5cm}
\date{\today}

\begin{abstract}
Under the condition of partial surface wettability, thin liquid films can be destabilized by small perturbations and rupture into droplets. As successfully predicted by the thin film equation (TFE), the rupture dynamics are dictated by the liquid-solid interaction. The theory describes the latter using the disjoining pressure or, equivalently, the contact angle. The introduction of a secondary fluid can lead to a richer phenomenology thanks to the presence of different fluid/surface interaction energies but has so far not been investigated. In this work, we study the rupture of liquid films with different heights immersed in a secondary fluid using a multi-component lattice Boltzmann (LB) approach. We investigate a wide range of surface interaction energies, equilibrium contact angles, and film thicknesses. We found that the rupture time can differ by about one order of magnitude for identical equilibrium contact angles but different surface free energies. Interestingly, the TFE describes the observed breakup dynamics qualitatively well, up to equilibrium contact angles as large as 130$^\circ$. A small film thickness is a much stricter requirement for the validity of the TFE, and agreement with LB results is found only for ratios $\epsilon=h/L$  of the film height $h$ and lateral system size $L$ such as $\epsilon\lesssim\times10^{-3}$.
\end{abstract}

\maketitle


\section{\label{sec:intro}Introduction}

Dewetting is the spontaneous, reverse process of wetting of a liquid spreading on a solid surface~\cite{deGennes85,Sharma96,Craster09,Leroy16,Nepomnyashchy21}. This phenomenon can be observed in everyday life, for example, when pouring oil on a cooking pan, covering a glass surface with a water film, or when a tear film wets our eyes~\cite{Bhamla16,Suja22}. Controlling the dewetting dynamics is key to several industrial applications, including printable photovoltaics~\cite{Robinson11,Ishihara16,Howard19}, or lubrication and coating processes~\cite{Ma11,Habibi16,Barroso19}. In general, dewetting dynamics takes place when a thin liquid film, in contact with a partially wettable surface,  ruptures into droplets. In the language of thermodynamics, the film reaches its equilibrium droplet shape because the latter is energetically favorable with respect to the flat interface~\cite{Oron97,Bertrand10,Nepomnyashchy21}. 

Dewetting can happen because of intrinsic or extrinsic rupture mechanisms. Extrinsic mechanisms include rupture due to surface heterogeneities or the presence of impurities on the surface~\cite{Xue11,Youssef16}. These extrinsic mechanisms are opposed to intrinsic, spinodal dewetting ones. Spinodal dewetting occurs in extremely thin liquid films that break up spontaneously due to interface perturbations or thermal fluctuations~\cite{Mitlin93,Fetzer07,Zitz21}.
Several experimental works studied the evolution of thin films on horizontal partially-wettable surfaces~\cite{Seemann01,Becker03,Gonzalez07,Edwards16}, chemically structured walls~\cite{Checco12}, films with toroidal shape~\cite{Edwards21} or surrounded by a second viscous phase~\cite{Edwards20}. Pulsed-laser-induced dewetting of metal alloys has also been investigated~\cite{Kondic09,Trice07,Wu10,Kondic20,Diez21}.
The most popular model to describe dewetting phenomena is the celebrated TFE~\cite{Oron97}, which describes the space-time evolution of the film height profile $h({\mbox{\textbf{x}}},t)$. In its most simplified form for the deterministic case (i.e., excluding thermal fluctuations) and no-slip boundary conditions, the TFE reads~\cite{Munch05,Rauscher08}:
\begin{figure*}[t!]
    \centering
    \includegraphics[width=.8\linewidth]{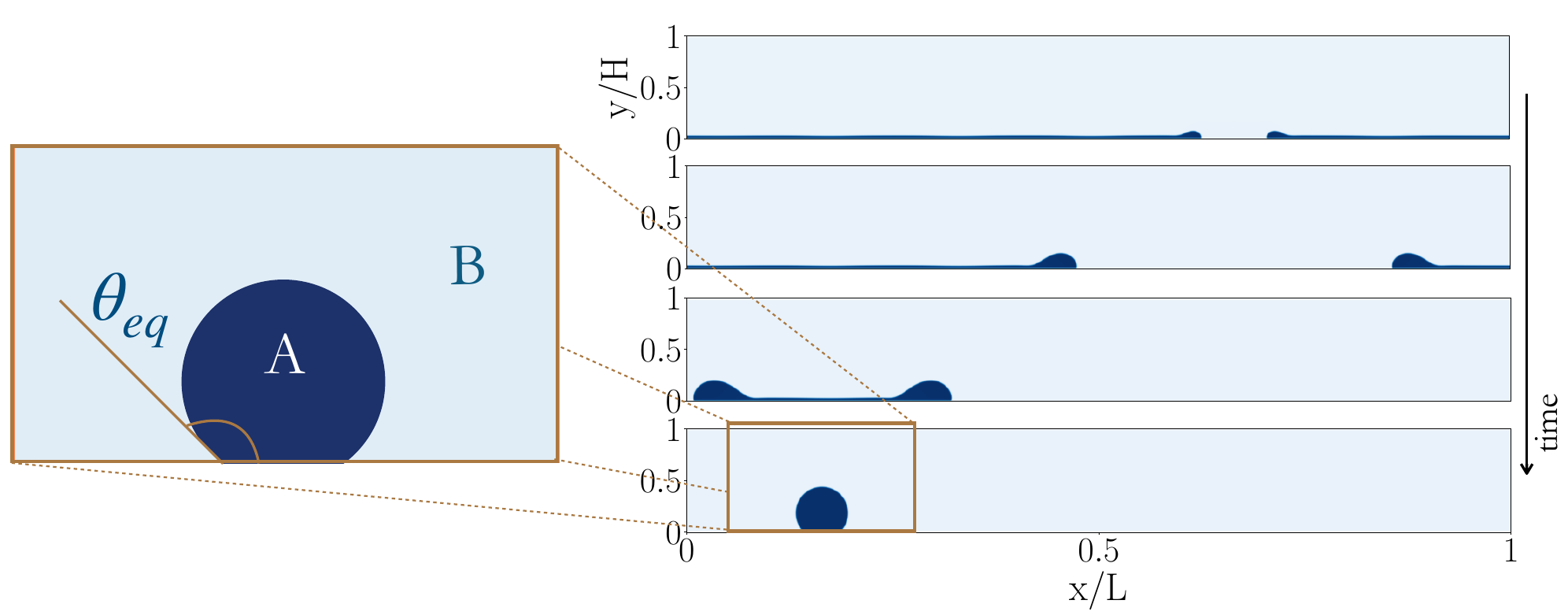}
    \caption{Sketch of the physical problem: the dewetting process takes place, causing the rupture of the thin liquid film (component A) and the transition to a droplet shape with a final contact angle $\theta_\mathrm{eq}$. Periodic boundary conditions are applied along the x-direction.\label{fig:sketch}}
\end{figure*}
\begin{equation}\label{eq:TFE}
 \partial_t h({\mbox{\textbf{x}}},t) = -\dfrac{1}{3 \mu}\nabla \cdot \left[ h({\mbox{\textbf{x}}},t)^3 \nabla \left( \Pi(h) + \gamma \nabla^2 h({\mbox{\textbf{x}}},t)\right) \right],
\end{equation}
where $\mu$ is the dynamic viscosity of the film, $\gamma$ its surface tension, and $\Pi(h)$ the disjoining pressure. The disjoining pressure describes the presence of interactions between the fluid and the solid surface, therefore incorporating information on the wetting properties, and is usually expressed in terms of the equilibrium contact angle $\theta_\mathrm{eq}$~\cite{Schwartz01}. The lubrication approximation~\cite{Joseph13film}, which underpins Eq.~\eqref{eq:TFE}, assumes the ratio between the characteristic film height $h$ and length $L$ to be very small ($\epsilon = h/L \ll 1$). Furthermore, Eq.~\eqref{eq:TFE} is valid under the assumptions of negligible inertial effects (implying a small Reynolds number) and small contact angles due to the inevitable presence, in the theoretical description, of a precursor film. Numerical solutions of the TFE have been obtained using various approaches including contact line solutions~\cite{Thiele03,Eggers04,Snoeijer06}, gradient dynamics models~\cite{Thiele13,Thiele16,Henkel21}, as well as LB-based methods~\cite{Zitz19,Zitz21}, by allowing to observe the dynamics of $h({\bm x},t)$ within the TFE limits. Other numerical methods have been employed to go beyond the lubrication approximation, including phase-field approaches~\cite{Jiang12,Ronsin21}, single-phase LB models~\cite{Biferale07,Edwards16,Wang17}, and volume-of-fluid methods~\cite{Mahady13,Mahady16,Afkhami18}. These approaches take into account inertial effects and control the wettability condition by selecting an equilibrium contact angle rather than introducing a disjoining pressure.

However, a comprehensive investigation of thin-film rupture in a wide range of film height, surface tension, and contact angles that can overcome the limitations of the TFE is still missing. Here, we perform multi-component LB simulations of the thin-film rupture and subsequent dewetting process as sketched in Fig.~\ref{fig:sketch}. This model distinctly handles the dynamics of the liquid film and the surrounding fluid, allowing us to explore a large number of different physical situations. Indeed, we systematically explore a wide range of parameters, reaching equilibrium contact angles close to $180^\circ$. As we will show, the predictions of the TFE in terms of the power spectrum of the film height are in surprisingly good agreement with our simulations for contact angles as high as $130^\circ$, and deviations from the TFE prediction are observed from a film thickness $\epsilon$ larger than $3.5\times10^{-3}$.

\section{\label{sec:method}Method}
LB simulations~\cite{Kruger17, Succi18} solve the Boltzmann transport equation on a lattice. In the long-wavelength limit, Navier-Stokes equations emerge from it as described by the Chapman-Enskog expansion of the discretized Boltzmann equation.
We perform simulations of a binary mixture with components labeled $A$ and $B$. The mesoscopic dynamics of each fluid component  $\sigma=A$,$B$ is described in terms of the probability distribution functions $f^i_{\sigma}({\bf x},t)$ of finding a fluid particle in a specific discrete lattice node {\bf x} and a discrete instant time $t$. The index $i$ refers to a set of discrete velocities ${\bm c}^i$, which allow the propagation of $f^i_{\sigma}$ on the lattice. Here, we employ a D3Q19 lattice, i.e, a set of 19 velocity vectors ($i = 0, \dots, 18$) on each node of a three-dimensional lattice. Because of the symmetry of the problem, we simulate a quasi-two dimensional system by restricting one dimension of the problem (z) to two lattice units. In the LB method, the dynamics of the $f^i_{\sigma}$ follows the discretized Boltzmann transport equation~\cite{Kruger17, Succi18}
\begin{equation}\label{eq:LBeq}
\small
f^i_{\sigma}({\bf x}+{\bm c}^i,t+1) - f^i_{\sigma}({\bf x},t) = - \dfrac{1}{\tau}\left[ f^i_{\sigma}({\bf x},t) -f^{i,(eq)}_{\sigma}({\bf x},t)\right].
\end{equation}
For the sake of simplicity, in Eq.~\eqref{eq:LBeq} and hereafter, we fix the lattice spacing $\Delta {\bm x}$ and the time step $\Delta t$ to one. The left-hand side of Eq.~\eqref{eq:LBeq} rules the streaming of $f^i_{\sigma}$ on the lattice, while the right-hand side represents the collision term. This operator models the relaxation of $f^i_{\sigma}$ towards the discretized local Maxwellian distribution $f^{i,(eq)}_{\sigma}({\bf x},t)$ with a relaxation time scale $\tau$.  The explicit shape of $f^{i,(eq)}_{\sigma}({\bf x},t)$ is given by~\cite{Kruger17} (repeated indeces are summed up)
\begin{equation}\label{eq:feq}
\small
f^{i,(eq)}_{\sigma}({\bf x},t) = w_i \rho_\sigma \left[1+\dfrac{u_{k,\sigma} c^i_{k}}{c^2_s} + \dfrac{u_{k,\sigma} u_{j,\sigma} (c^i_{k} c^i_{j} - c^2_s \delta_{kj})}{2 c^2_s} \right],
\end{equation}
where $c_s=1/\sqrt{3}$ is the speed of sound and $w_i$ are lattice-dependent weights, which, for the D3Q19 lattice, are $w_i = 1/3$ for $i =0$, $w_i = 1/18$ for $i =1\ldots6$, $w_i = 1/36$ for $i =7\ldots18$, respectively. 
The fluid component densities $\rho_\sigma$, the total density $\rho$ and the momentum $\rho {\bm u}$ can be computed from the populations as $\rho_\sigma = \sum_i f^i_{\sigma}({\bf x},t)$, $\rho = \sum_\sigma \rho_\sigma$ and  $\rho {\bm u}({\bf x},t) = \sum^i_{\sigma} {\bm c}^i f^i_{\sigma}({\bf x},t)$, while the dynamic viscosity $\mu$ follows the LB relation $\mu_\sigma = \rho_\sigma c_s^2(\tau-1/2)$.

In order to observe phase separation between the two components, it is necessary to include fluid-fluid interactions~\cite{Liu16}. We employ the model proposed by Shan and Chen (SC)~\cite{ShanChen93,ShanDoolen95}, where a force ${\bm F}_\sigma ({\bf x},t)$ acts on component $\sigma$, entering implicitly in Eq.~\eqref{eq:feq} through a shift in the definition of the momentum:
\begin{equation}
{\bm u}_{\sigma} ({\bf x},t)= {\bm u}({\bf x},t) + \dfrac{\tau {\bm F}_\sigma ({\bf x},t)}{\rho_\sigma}.
\end{equation}

The term ${\bm F}_\sigma ({\bf x},t)$ contains both fluid-fluid as well as wall-fluid interactions. In the SC model, the fluid-fluid interaction term ${\bm F}^{ff}_\sigma$ takes the form 
\begin{equation}\label{eq:SC}
{\bm F}^{ff}_\sigma ({\bf x},t) = - \mbox{G}_{AB} \psi_\sigma ({\bf x},t) \sum_i w_i \psi_{\sigma '}({\bf x}+{\bm c}^i,t) {\bm c}^i,
\end{equation}
where $\sigma$ and $\sigma'$ (with $\sigma\ne\sigma'$) denote the two components and  $G_{AB}>0$ tunes the (repulsive) interaction strength. Here, the so-called pseudopotential $\psi_\sigma ({\bf x},t)$ coincides with the fluid-component density $\psi_\sigma ({\bf x},t) = \rho_\sigma ({\bf x},t)$. Notice that the implementation of the SC-LB yields a diffuse interface between fluid components with thickness $w_{int}$.

We introduce the wall-fluid interaction following the approach of Huang and coworkers~\cite{HuangSukop07}. We define the pseudo-potential for the solid wall as
\begin{equation}
s ({\bf x}) = \begin{cases}
1 \ \ \ \ \ \ \ {\mbox{\textbf{x}}} \in \mbox{wall}\\
0  \ \ \ \ \ \ \ {\mbox{\textbf{x}}} \in \mbox{fluid},
\end{cases}
\end{equation}
and, as a consequence, we can write the wall-fluid interactions as
\begin{equation}\label{eq:wallForcing}
{\bm F}^{wf}_\sigma ({\bf x},t) = - G_{\mbox{\tiny W}\sigma} \psi_\sigma \sum_i w_i s({\bf x}+{\bm c}^i) {\bm c}^i.
\end{equation}
Eq.~\ref{eq:wallForcing} can model either a repulsive ($G_{\mbox{\tiny W}\sigma} > 0$) or an attractive ($G_{\mbox{\tiny W}\sigma} < 0$) interaction. To prevent spurious forces generated by the presence of a strong gradient of the components $A$ and $B$, we set the value of $\psi_\sigma$ in each wall node to that of the opposite fluid node. 

The simulation box is periodic in the x- and z- directions, while the fluid is enclosed between two walls along the (vertical) y-direction. A half-node bounce-back rule implements second-order no-slip boundary conditions at the walls~\cite{Kruger17}. 
Hereafter we refer to the liquid film as the $A$ component, and we report all quantities in lattice units (lbu).

We choose the following set of parameters and initial conditions to study the film rupture: the initial film profile along y for fluid $A$ is a step function of width $h_0$ in the range from 6 to 13 lbu, which drops from $\rho_A=1$  (for y$\le h_0$) to $\rho_A=0.027$ (for y$> h_0$). Conversely, the density of fluid $B$ raises from $\rho_B=0.027$ (for y$\le h_0$) to $\rho_B=1$ (for y$>h_0$). This choice of density values sets a viscosity ratio $\mu_A/\mu_B = 1$, with $\tau$ being kept fixed to 1 in all simulations. The interaction parameter $G_{AB}$ ranges from 1.4 (the minimum value ensuring phase separation) to 1.7, above which the film is always stable and does not rupture. The range of $G_{AB}$ is directly related to the choice of $\rho_A$ and $\rho_B$, as highlighted by the corresponding phase-separation diagram (see Supplementary Material, Fig. S1(a)). In the absence of the wall, these values of $G_{AB}$ lead to interfacial widths in the range from 2 to 4 lbu (see Supplementary Material, Fig. S1(b)). To ensure the stability of the simulation, we use wall-fluid interaction parameters $G_{WA}$ and $G_{WB}$ in the range from $-0.4$ to  $0.4$. $G_{AB}$ is directly related to the surface tension $\gamma$~\cite{ShanChen93}, which can be measured via Laplace experiments (see Supplementary Material, Fig. S1(b)). The definition of the wall-fluid interactions in Eq.~\eqref{eq:wallForcing} suggests that $G_{WA}$ and $G_{WB}$ are related to the wall-fluid interfacial tensions $\gamma_{WA}$ and $\gamma_{WB}$, which, however, cannot be measured directly. Instead, for a given value of $\gamma$, their difference  $\Delta G_W = G_{WB}-G_{WA}$ tunes the value of the equilibrium contact angle $\theta_\mathrm{eq}$. The latter is expected to be proportional to $\Delta G_W$, as observed by Huang and coworkers~\cite{HuangSukop07}. Hereafter, all dimensional quantities will be reported in lattice Boltzmann units (lbu). All simulations are performed on a domain of size $L = 2048$ lbu along the x-direction and $H = 256$ lbu along the vertical y-direction.
\begin{figure}[t!]
\centering
\includegraphics[width=1.\columnwidth]{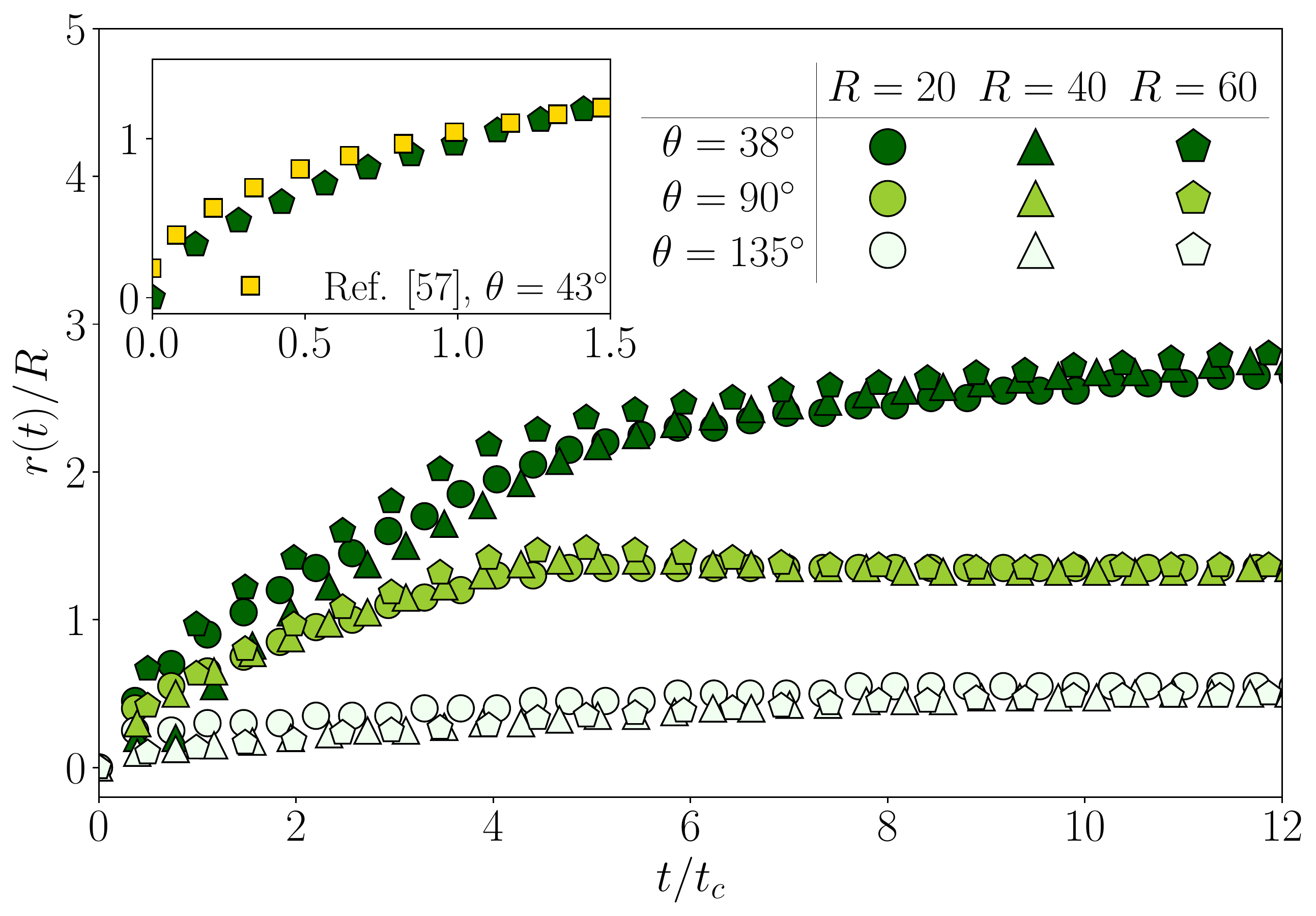}
\caption{Time evolution of the droplet contact area radius $r$, rescaled by the droplet initial radius $R$, as a function of the time, rescaled by the characteristic wetting time $t_c = (\rho R^3/\gamma)^{1/2}$. Different values of $R$ and contact angle $\theta$ are represented with different symbols and colors, respectively. The inset shows the agreement with experimental data~\cite{Bird08} corresponding to $\theta = 43^\circ$.\label{fig:BirdLaw}} 
\end{figure}
\begin{figure}[t!]
    \centering
    \includegraphics[width=.85\columnwidth]{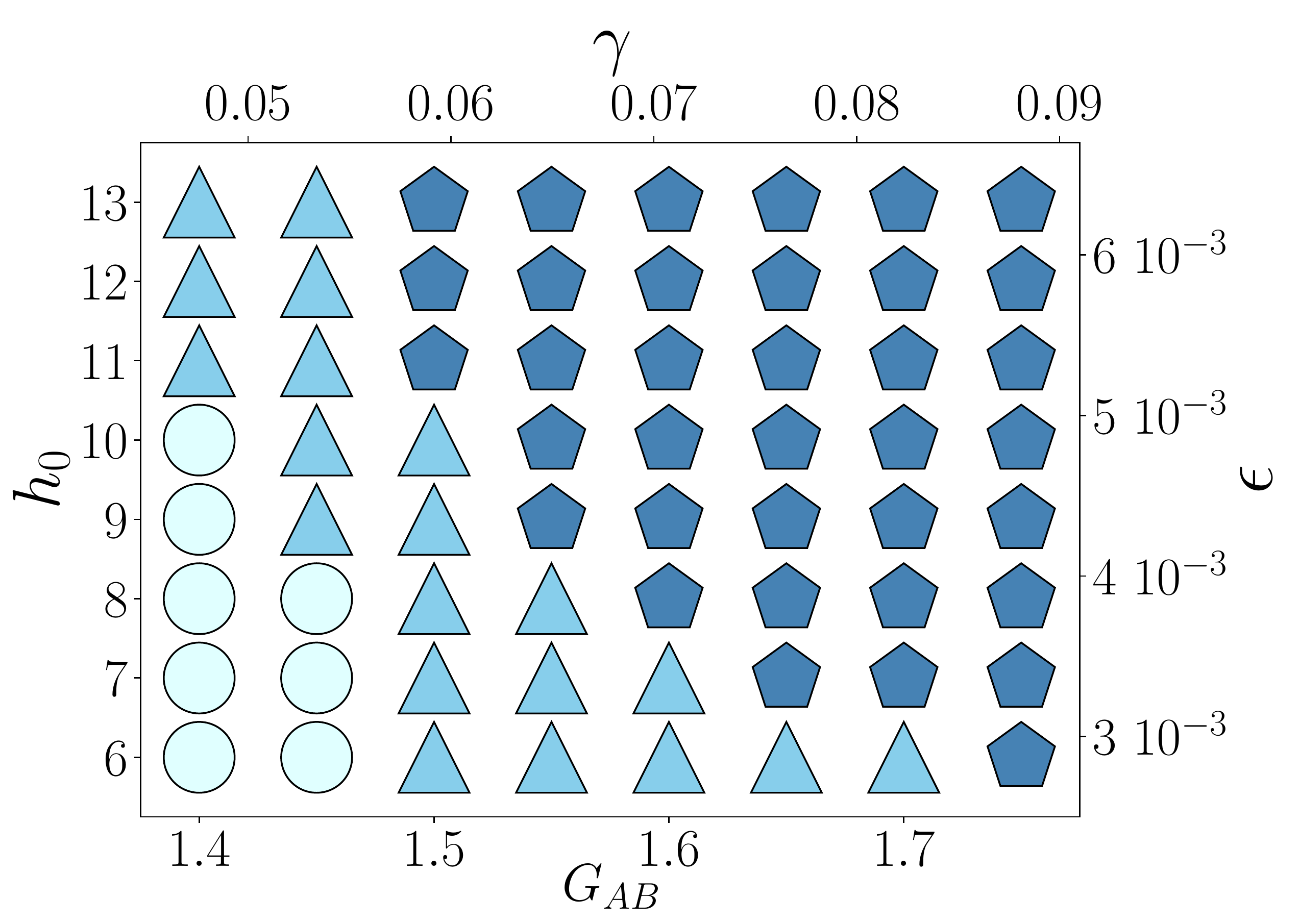}
    \caption{Stability diagram of thin liquid films as a function of initial nominal height $h_0$ and fluid-fluid interaction strength $G_{AB}$. The corresponding values of $\epsilon = h_0/L$ and the surface tension $\gamma$, respectively, are also shown. Three main regions are observed: (1) stability region (pentagons)
    , (2) conditional stability region (triangles) 
    , and mixing region (circles).
    All dimensional quantities are reported in lbu. Further details can be found in the text.\label{fig:stability}}
\end{figure}

To validate our implementation of the pseudopotential lattice Boltzmann model by Shan and Chen, we perform simulations of the wetting/spreading dynamics of a liquid droplet on a flat wall. We measure the contact angle as a function of the wall-fluid interaction parameters $G_{WA}$ and $G_{WB}$, recovering the estimates reported by Sukop and coworkers~\cite{HuangSukop07}. In addition, we study the droplet-spreading problem by replicating the experimental approach of Bird and coworkers~\cite{Bird08}, who measured the temporal evolution of the contact area radius $r(t)$. By varying the initial radius $R$ and the fluid-wall interactions, they observed a data collapse on several master curves when properly rescaling the time with respect to the characteristic wetting time $t_{c} = (\rho R^3/\gamma)^{1/2}$ and $r(t)$ to the initial value. Here, $\rho$ is the density of the liquid droplet. This result underlines that the wetting dynamics depends only on the fluid-wall interactions and not on $R$. Fig.~\ref{fig:BirdLaw} shows our SC-LB results, further confirming the validity of our implementation.
\section{\label{sec:results}Results and discussion}
\begin{figure*}[t]
    \centering
     \includegraphics[width=.9\linewidth]{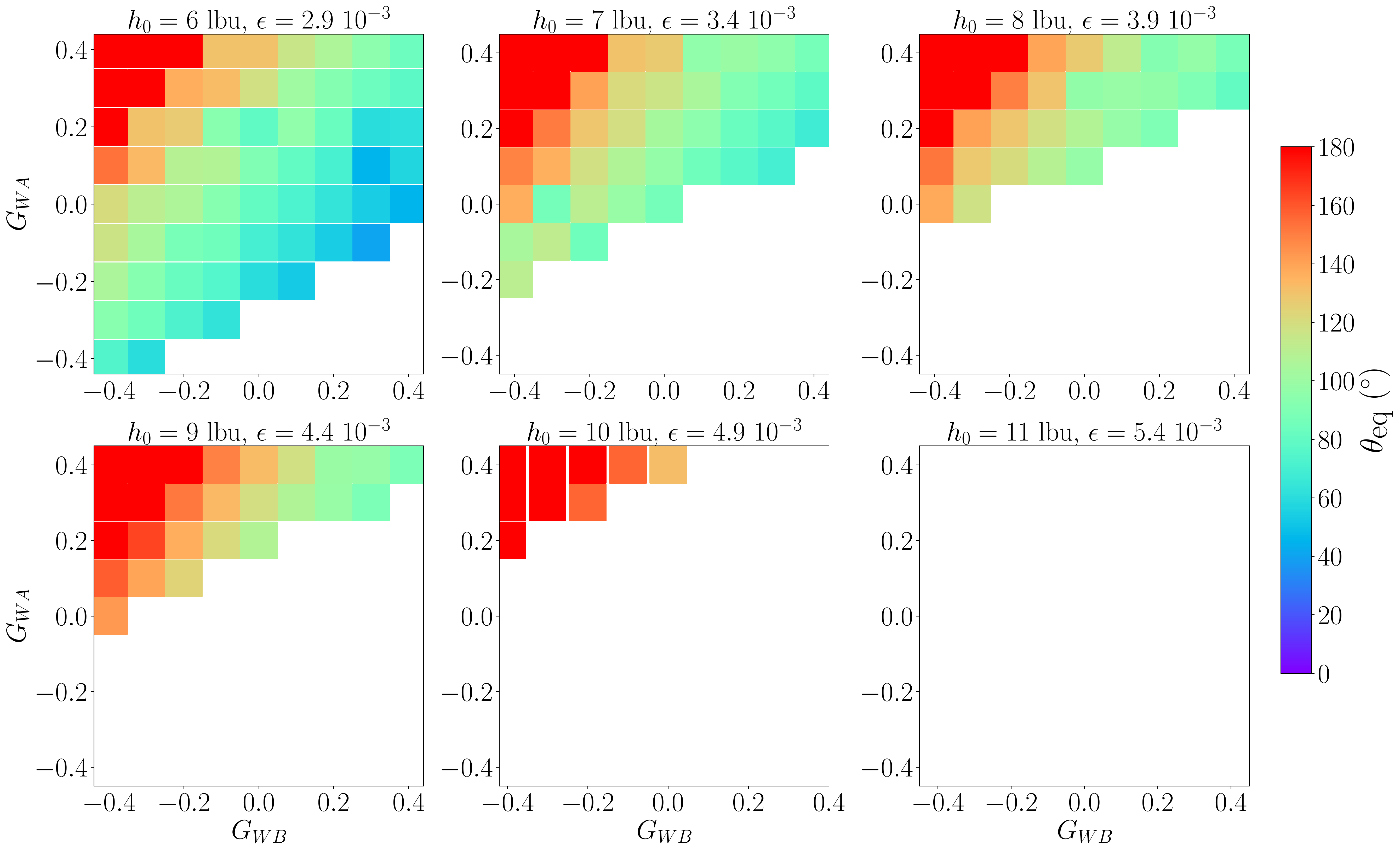}
    \caption{Equilibrium contact angle $\theta_\mathrm{eq}$ as a function of the fluid-wall interactions $G_{WA}$ and $G_{WB}$ corresponding to a vertical cut of Fig.~\ref{fig:stability}, i.e., values are reported at fixed fluid-fluid interaction strength $G_{AB} = 1.5$ ($\gamma = 0.059$ lbu) and different initial nominal height $h_0$ (and $\epsilon$). White regions refer to the condition in which the film is stable upon perturbation.\label{fig:contactAngles}}
\end{figure*}
\begin{figure*}[t!]
    \centering
  \includegraphics[width=.9\linewidth]{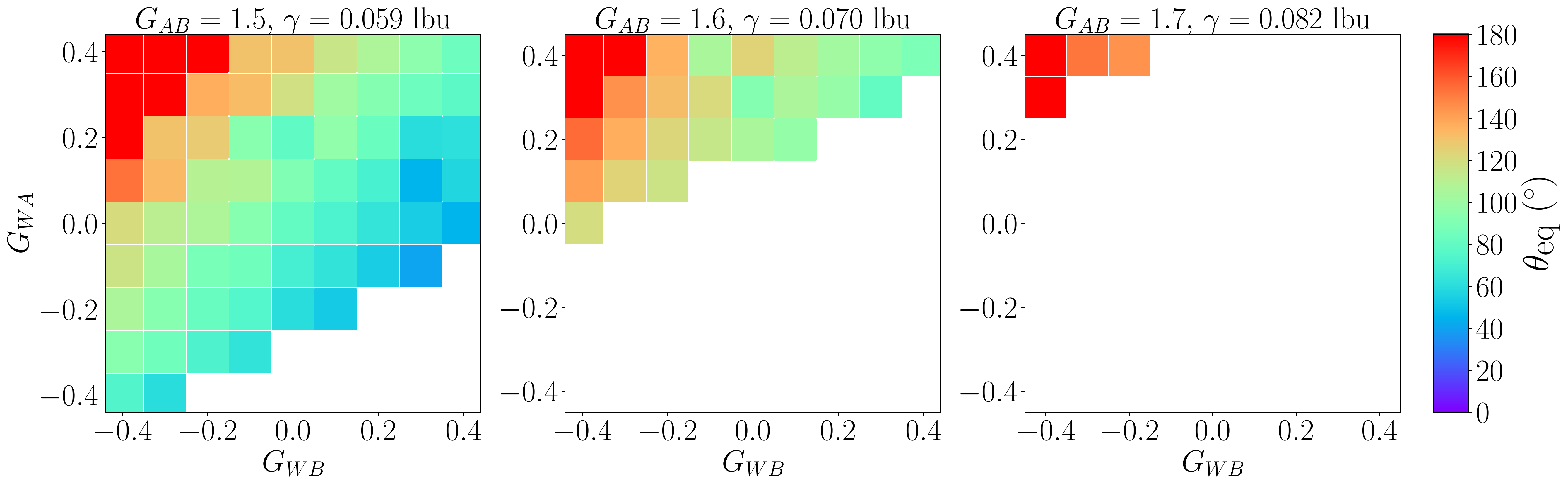} \\
     \caption{Equilibrium contact angle $\theta_\mathrm{eq}$ as a function of the fluid-wall interactions $G_{WA}$ and $G_{WB}$ corresponding to a horizontal cut of Fig.~\ref{fig:stability}, i.e., values are reported at fixed initial nominal height $h_0 = 6$ lbu ($\epsilon = 2.9 \ 10^{-3}$) and different coupling parameter $G_{AB}$ (i.e., surface tension $\gamma$). White regions refer to the condition in which the film is stable upon perturbation.\label{fig:contactAngles2}}
\end{figure*}
\begin{figure*}[t!]
\centering
\includegraphics[width=.9\linewidth]{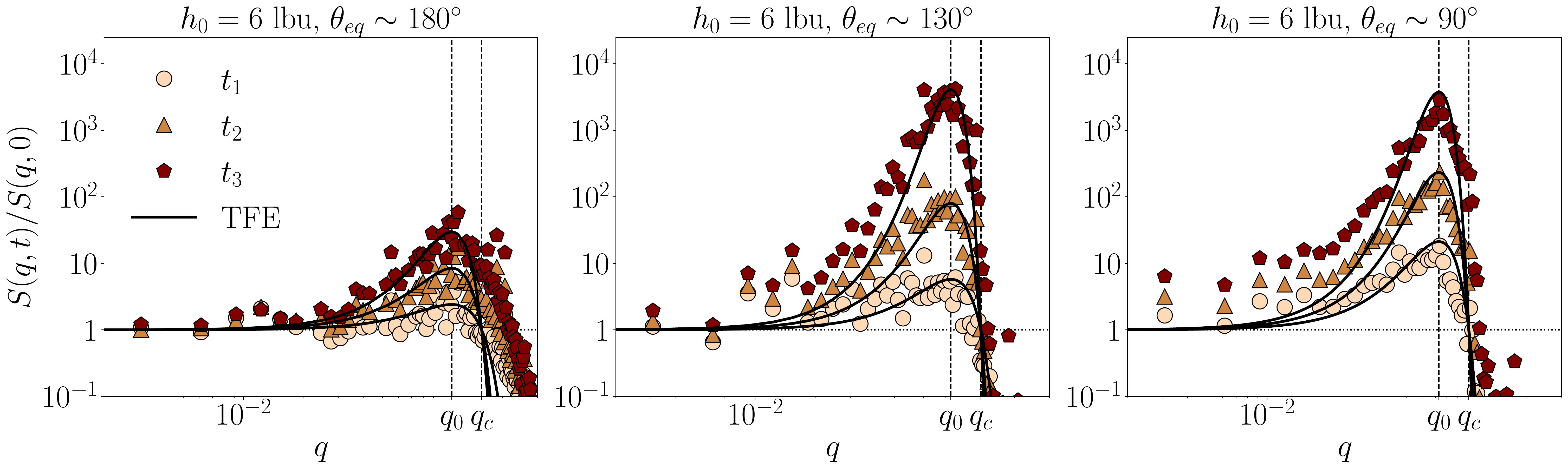}\\
\includegraphics[width=.9\linewidth]{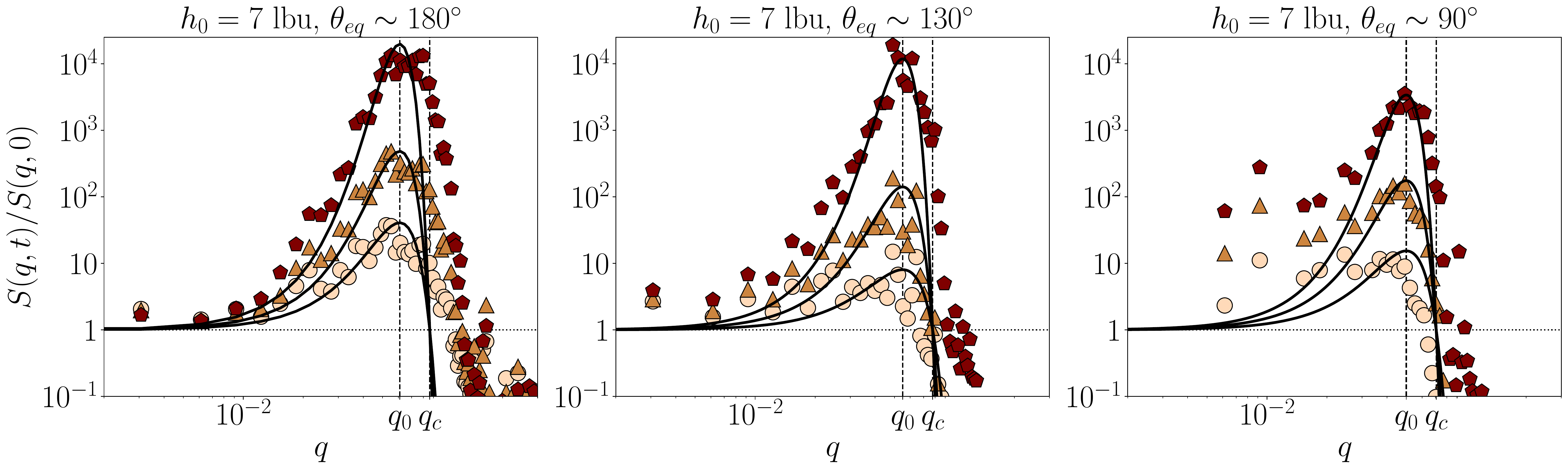}\\
 \includegraphics[width=.9\linewidth]{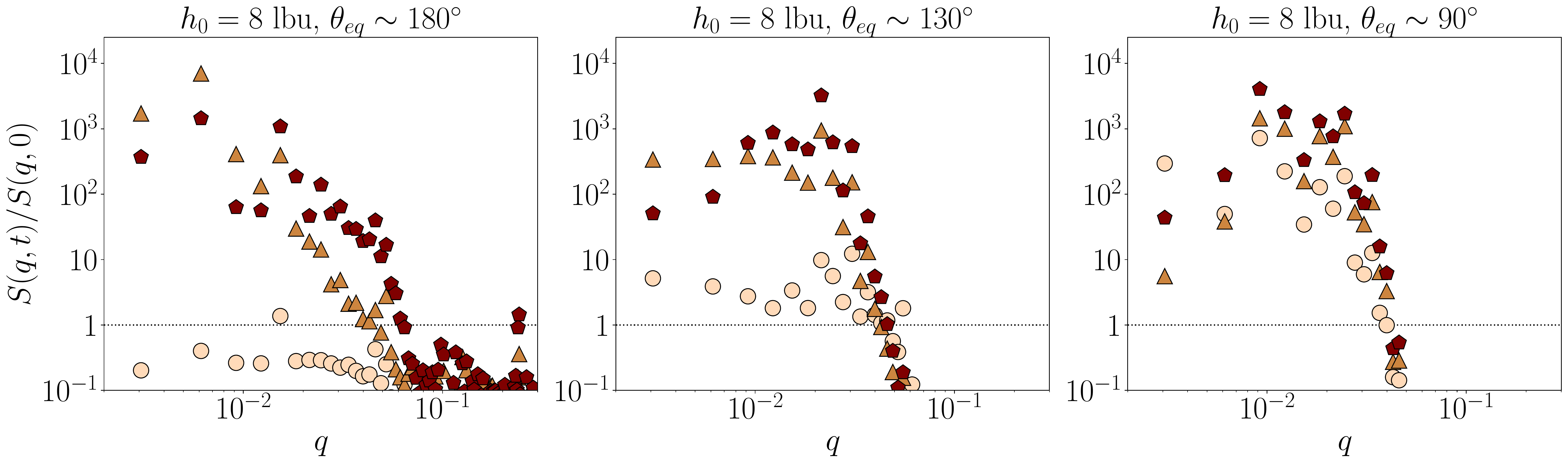}
\caption{Power spectra $S(q,t)$, normalised to the initial one $S(q,0)$ for three instants of time (different colours/symbols), different $h_0$ and $\theta_\mathrm{eq}$. For cases with $h_0 \le 7$ lbu, a comparison with the TFE predictions is also shown (solid black lines). Vertical dashed lines refer to the fastest ($q_0$) and critical ($q_c = \sqrt{2}q_0$) growing modes. All panels have the same q-range. The time values $t_1$,$t_2$ and $t_3$ are reported in the Supplementary Material.\label{fig:TFE}}
\end{figure*}

We simulate the dewetting dynamics of a thin liquid film placed on a flat wall and the resulting film rupture (see Fig.~\ref{fig:sketch}) in the following way: we start with a uniform film of initial, nominal height $h_0$; we perform a dedicated simulation run to relax this initial condition: due to the diffuse nature of the SC-LB interface, the film height quickly reaches the equilibrium initial (i.e., before the perturbation) height $h_\mathrm{initial}$, calculated as the distance from the wall at which the density $\rho_A(\mbox{y})$ reaches a threshold density $\rho_\mathrm{th}$ set to half of its maximum value. Notice that in the absence of perturbations, because of the enforced translational invariance, the system is always stable and can reach the equilibrium density distribution along the normal direction without breaking. After this relaxation stage, we perturb the film with a random perturbation of its density in its interfacial region with width $w_{int}$. The perturbed density then reads
\begin{equation}
  \rho'_A({\bf x}, t) = \rho_A ({\bf x}, t) [1+P\beta] \hspace{1cm} {\bf x}\in w_{int},
\end{equation}
where the perturbation amplitude $P=10^{-3}$, and $\beta$ is a random variable uniformly distributed in [-1,1]. In this way, the film height is also perturbed as $h'(\mbox{x}, t) = h_\mathrm{initial} + \delta h(\mbox{x}, t)$.  We explicitly check that $P$ is small enough and that the results do not depend on its actual value. 

By considering all these ingredients together, we are ready to study the stability conditions for thin liquid films as a function of $G_{AB}$ (i.e., the surface tension $\gamma$) and $G_{WA}$ and $G_{WB}$ (i.e., the fluid-wall interactions). We compute stability diagrams by checking the conditions under which the film is stable or ruptures after the initial perturbation. 
In Fig.~\ref{fig:stability} we show the stability diagram as a function of $h_0$ (and the corresponding ratio $\epsilon = h_0/L$) and $G_{AB}$ (with the corresponding surface tension $\gamma$). We can distinguish three main regions: pentagons refer to a stable film under all fluid-wall interactions, triangles to conditional stability, for which at least one choice of the fluid-wall interactions triggers the film rupture, and circles for cases where the surface tension is not strong enough to preserve the phase separation. In the latter case, the two components (which would demix in the absence of the wall) start mixing.
\begin{figure*}[th!]
    \centering
    \includegraphics[width=.33\linewidth]{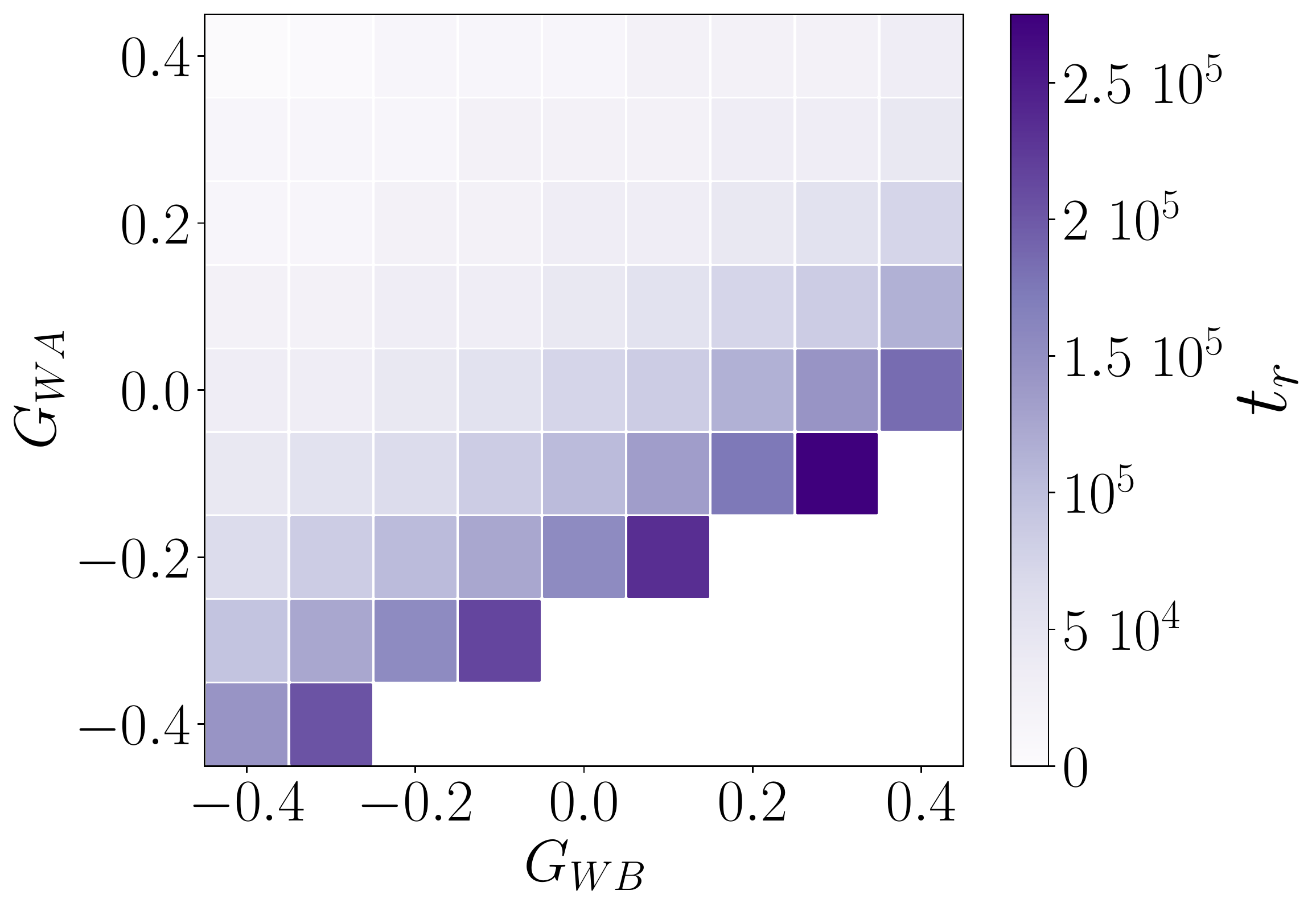}\includegraphics[width=.63\linewidth]{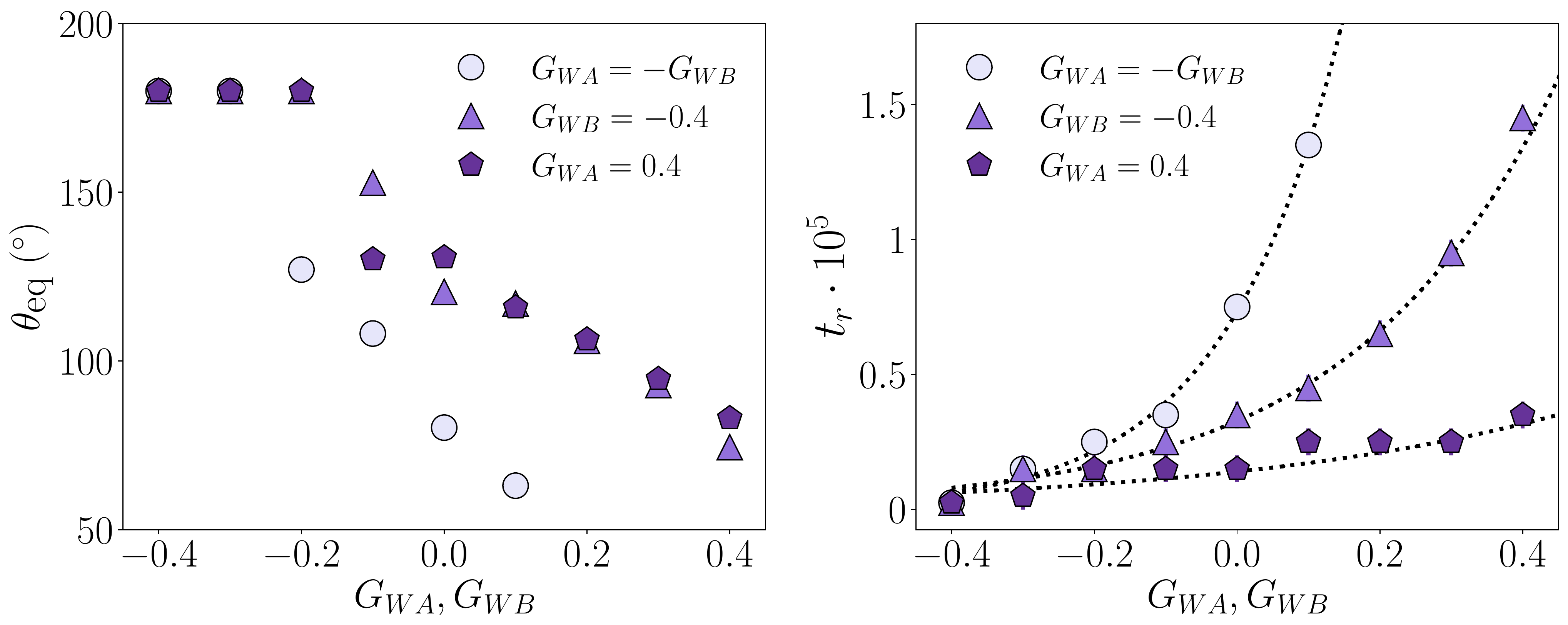}
    \begin{tabular}{m{0.4\textwidth}m{0.3\textwidth}m{0.3\textwidth}}
            \center \small (a) & \center \small (b) & \center \small (c)
    \end{tabular}
    \caption{Measure of the rupture time $t_r$ for a thin film with nominal initial height $h_0 = 6$ lbu (i.e., $\epsilon = 2.9 \ 10^{-3}$), and fluid-fluid coupling $G_{AB} = 1.5$ (i.e., $\gamma = 0.059$ lbu). Panel (a): map of $t_r$ for different fluid-wall interactions (corresponding to Fig.~\ref{fig:contactAngles2}, left panel). Panel (b): equilibrium contact angle $\theta_\mathrm{eq}$ as a function of the coupling parameters extracted from three different cuts of plot in panel (a). Circles correspond to a diagonal cut, i.e., by fixing $G_{WA} = -G_{WB}$; triangles correspond to a vertical cut at $G_{WB} = -0.4$; pentagons correspond to a horizontal cut at $G_{WA} = 0.4$. We report this panel in order to highlight that we are equivalently moving from $\theta_\mathrm{eq} = 180^\circ$ to $\theta_\mathrm{eq} = 90^\circ$ along the vertical and horizontal cuts. Panel (c): $t_r$ as a function of a wall-fluid coupling parameter for the three cuts. Dotted lines refer to exponential functions that we draw to facilitate the reader's eye in distinguish the different behaviour of $t_r$ as a function of the fluid-wall interactions. All cuts are shown at decreasing equilibrium contact angles (see panel (b)). Symbols and colours correspond to those in panel (b).\label{fig:ruptureTime}}
\end{figure*}

The cases of conditional stability require to explore further the effect of the fluid-wall interactions. $G_{WA}$ and $G_{WB}$ affect the stability in a complex way due to the simultaneous influence of $h_0$ and $G_{AB}$. In Figs.~\ref{fig:contactAngles} and~\ref{fig:contactAngles2} we show the conditional stability region (and the equilibrium contact angle $\theta_\mathrm{eq}$ for the unstable cases) along two cuts. One cut is performed at variable $h_0$ and fixed $G_{AB} = 1.5$ (i.e., $\gamma = 0.059$ lbu) and the other one is performed at variable $G_{AB}$ and fixed $h_0 = 6$ lbu. In all panels of Figs.~\ref{fig:contactAngles} and~\ref{fig:contactAngles2}, the white areas refer to the condition in which the film is stable upon perturbation. 

An increase of $h_0$ at fixed $G_{AB}$ and an increase of  $G_{AB}$ at fixed $h_0$ both result in overall improved stability of the film. The general trend does not come as unexpected. However, while $G_{WA}$ and $G_{WB}$ play a symmetric role in the determination of the contact angle, as already noticed by Sukop and coworkers~\cite{HuangSukop07}, this is not true anymore for the determination of the stability region. This result highlights the role played by each of the components in the dewetting dynamics: the best conditions to induce the transition are those for which the substrate repels one component (the film) while the other one is attracted by it. The TFE, which is valid for a single fluid only, turns out that the film stability depends only on the equilibrium contact angle $\theta_\mathrm{eq}$ (appearing in the definition of the disjoining pressure~\cite{Schwartz01}), that is, on $\Delta G_W$,~\cite{Munch05,Rauscher08,HuangSukop07}. Based on these considerations, SC-LB simulations appear to be useful for simulating dewetting dynamics by providing a much richer phenomenology. The importance of considering two components playing a role in the dewetting dynamics will also be confirmed later with the measurement of rupture times.

After clarifying these points, we are now in the condition to investigate the dewetting process at large contact angles and check the validity of the TFE beyond its original limits. This step has a fundamental importance in stressing the solid foundations as well as the strength of the implemented numerical method. The time-dependent structure factor of the height profile provides an insightful route to this aim. Starting from the  perturbation growth $\delta h(\mbox{x}, t) = h_\mathrm{initial} - h(\mbox{x},t)$ and its Fourier transform
\begin{equation}\label{eq:perturbH}
\delta \hat{h}(q,t) = \frac{1}{\sqrt{2\pi}}\int^\infty_{-\infty} \delta h(\mbox{x}, t) e^{-iq\mbox{x}} \ d\mbox{x},
\end{equation}
one can define the structure factor (i.e., the power spectrum of the perturbation) as 
\begin{equation}
    S(q,t) = |\delta \hat{h}(q,t)|^2,
\end{equation}
where $q$ is the wave number. In Fig.~\ref{fig:TFE} we show $S(q,t)$ measured for three different heights ($h_0=6$, 7 and 8 lbu) and three different equilibrium contact angles $\theta_\mathrm{eq}\ge90^\circ$, normalized to the initial value $S(q,0)$. 
The reported values are the result of averages over about 100 independent runs, taken at three different times $t_1$, $t_2$ and $t_3$, with $t_1$ corresponding to an early time after the perturbation (but with $t_1$ large enough for $S(q,t_1)/S(q,0)$ to be appreciably different from 1), $t_3$ to a time near the rupture, and $t_2$ an intermediate time. These times are different for each panel, and are reported in the Supplementary Material, Tab.~S1.
The structure factors $h_0=6$ lbu have the qualitative features expected from the TFE as they show a maximum (defining the fastest growing mode $q=q_{0}$) followed by a steep descent towards 1 in $q=q_{c}$, the so-called critical growth mode. At values larger than $q_c$ the structure factor is smaller than one, implying that the modes with $q>q_c$ decrease in amplitude with time and are thus stable. Modes with $q<q_c$ grow indefinitely and are the unstable ones that lead to the film rupture. For values of the contact angle smaller than about $130^\circ$, $S(q,t)$ does not only agree qualitatively with the prediction of the TFE, but also quantitatively. The solid black lines reported in Fig.~\ref{fig:TFE} show the prediction of the deterministic TFE  $S(q,t)/S(q,0) =  e^{\omega(q)t}$, where the dispersion relation $\omega(q)$ is given by~\cite{Rauscher08}
\begin{equation}\label{eq:dispersion}
    \omega(q) = \dfrac{\gamma h_\mathrm{eff}^3 q_0^4}{3\mu}\left[2\left(\frac{q}{q_0} \right)^2 -  \left(\frac{q}{q_0} \right)^4\right].
\end{equation}
\begin{table}[]
\begin{tabular}{m{0.082\textwidth}m{0.082\textwidth}m{0.082\textwidth}m{0.082\textwidth}m{0.082\textwidth}}
$h_0$ (lbu) & $\theta_\mathrm{eq}$ ($^{\circ}$)& $h_\mathrm{initial}$ (lbu) & $h_\mathrm{eff}$ (lbu)& $q_0$ (lbu)\\
\hline
\hline
6 & 180 & 5.23 & 5.25 & 0.111\\
6 & 130 & 3.79 & 4.07 & 0.096\\
6 & 90 & 3.69 & 4.57 & 0.073\\
7 & 180 & 7.31 & 8.81 & 0.061\\
7 & 130 & 4.91 & 5.85 & 0.055\\
7 & 90 & 4.77 & 5.16 & 0.050\\
\hline
\end{tabular}
\caption{Values of the measured initial height $h_\mathrm{initial}$, the effective height $h_\mathrm{eff}$ entering in Eq.~\eqref{eq:dispersion} and the fastest-growing mode $q_0$ for data shown in top panels ($h_0 = 6,7$ lbu) of Fig.~\ref{fig:TFE}.\label{tab:2}}
\end{table}
Here, the fitting parameter $h_\mathrm{eff}$ is an effective initial film height, which does not precisely coincide with the value of $h_\mathrm{initial}$. We remind that $h_0$ is the width of the initial step-like density distribution used to set up the simulation. This initial distribution relaxes in absence of perturbation to yield a diffuse interface whose half-maximum value is located at $h_\mathrm{initial}$. The choice of half-maximum (rather than another value) is somewhat arbitrary. In fact, using an effective width $h_\mathrm{eff}$ is necessary to compare our diffuse-interface simulation results with the TFE's sharp-interface ones. The values are reported in Tab.~\ref{tab:2}, where one can see that $h_\mathrm{eff}$ is systematically larger than $h_\mathrm{initial}$ in a range from 1 to 25\%. The only free fitting parameter used here is the fastest-growing mode value $q_0$, which is not directly modeled within the LB approach. 
In the deterministic TFE, $q_0$ takes the form
\begin{equation}
    q_0 = \left[ \frac{\partial\Pi}{\partial h_\mathrm{eff}}\frac{1}{2\gamma}\right]^{1/2},
\end{equation}
showing that increasing the surface tension while keeping the fluid-wall interaction constant should lead to an increased film stability. The vertical dashed lines in Fig.~\ref{fig:TFE} show the values of $q_0$ and the critical growing mode $q_c$ following the TFE predictions ($q_c = q_0\sqrt{2}$, with unstable modes being present for $0<q<q_c$). 
The match is superior when both $h_0$ and $\theta_\mathrm{eq}$ are small (top-right panel of Fig.~\ref{fig:TFE}), as the assumptions underlying the TFE are more accurately satisfied. Even in these cases, however, there are low-$q$ tails that are not well reproduced by the TFE. These tails could be the effect of the LB method's finite (albeit low) compressibility, as seen from the condition $S(q,t)>S(q,0)$ for low values of $q$, which implies an enlargement of the overall width of the film at time $t>0$. When $\theta_\mathrm{eq} = 180^{\circ}$, the data depart from the TFE prediction that $q_c = q_0\sqrt{2}$. 
The agreement achieved for $h_0=6$ deteriorates at $h_0=7$ and it is completely lost at $h_0\ge8$ (that is, $\epsilon > 3.5\times{}10^{-3}$), where a more complex behavior emerges, which includes also a time dependency of $q_c$ for $\theta_\mathrm{eq}=180^\circ$. 

The dewetting dynamics can be strikingly different even if the equilibrium contact angle is the same for two different fluid-wall interactions.
In Fig.~\ref{fig:ruptureTime}(a) we report the rupture time $t_r$ as a function of the fluid-wall interactions for the selected case $h_0 = 6$ lbu and $G_{AB} = 1.5$ (i.e., $\gamma = 0.059$ lbu) (see Fig.~\ref{fig:contactAngles2}, left panel).
We define $t_r$ as the time when the film density close to the wall becomes lower than the threshold $\rho_\mathrm{th}$  (half of the maximum density) at least at one lattice node.
By comparing the left panel of Fig.~\ref{fig:contactAngles2} with Fig.~\ref{fig:ruptureTime}(a), we note that fluid-wall interactions driving to the same $\theta_\mathrm{eq}$ lead to different rupture times.
More in detail, we report in Fig.~\ref{fig:ruptureTime}(c) $t_r$ as a function of one of the two wall-fluid interaction strengths $G_{WA}$ or $G_{WB}$ for three different cuts of panel (a), namely:  a diagonal cut (circles, $G_{WA}=-G_{WB}$); a vertical cut (triangles, $G_{WB}=-0.4$); and, a horizontal cut (pentagons, $G_{WA} = 0.4$).
For all cases, $t_r$ follows a behavior as a function of the wetting parameter (dotted lines in Fig.\ref{fig:ruptureTime}(c)) that we fit with an exponential function as a visual aid. The growth can be dramatically different depending on the wetting parameter, even when the equilibrium contact angle is the same, as shown in Fig.~\ref{fig:ruptureTime}(b). The plateau observed in Fig.~\ref{fig:ruptureTime}(b) for large contact angles is expected because as soon as the condition of super-hydrophobicity is reached, an increase of the fluid-wall interaction strength leaves the system unaltered. Furthermore, the measure of $\theta_\mathrm{eq}$ is affected by an error caused by the presence of the diffuse interface that explains the observed jump of $\theta_\mathrm{eq}$ next to the plateau. This picture confirms that the dynamics is not simply driven by $\theta_\mathrm{eq}$, but by a more complex combination of the two wall-fluid interactions.
If the film is subject to a strong repulsion ($G_{WA} = 0.4$), the other component has a less prominent effect. On the contrary, if the other component is strongly attracted to the wall ($G_{WB} = -0.4$), a slight variation in the film-wall interaction contributes to a substantial change in the rupture dynamics.

\section{Conclusions}\label{sec:conclusions}

We performed numerical simulations of the stability and rupture dynamics of liquid films on flat solid surfaces, immersed in a secondary fluid, using Shan-Chen multi-component lattice Boltzmann simulations.
We characterized the stability conditions of the films in terms of initial film height, surface tension, and equilibrium contact angle, spanning a wide range of these parameters up to full dewetting.
The functional form of the structure factor provided by the TFE turns out to be valid also for the two-component system up to angles of $130^\circ$, as long as the film aspect ratio $\epsilon=h/L$ is smaller than about $3.5\times10^{-3}$. The latter is the actual value of the aspect ratio that yields consistent results with the lubrication theory limit. Our analysis of the stability and rupture times underlines the richness in the phenomenology brought in by the presence of two liquid components. Another essential factor to consider is the viscosity ratio of the fluids, which we did not vary in the present investigation, but which is likely to influence the breakup dynamics. This variety opens the possibility of tuning the dynamics of dewetting over a wide range of time scales by controlling the fluid-wall interaction of the two fluids separately.

\section*{Supplementary Material}
Phase diagram, surface tension $\gamma$, and interface width $w_{int}$ as a function of $G_{AB}$, and list of $t_1$,$t_2$, and $t_3$ used to sample the data presented in Fig.~\ref{fig:TFE}.

\section*{Author declarations}
The authors have no conflicts to disclose.

\begin{acknowledgments}
This work has received financial support from the Deutsche Forschungsgemeinschaft (DFG, German Research Foundation) – Project-ID 431791331 – SFB 1452.
Furthermore, the authors acknowledge financial support by the DFG within the priority program SPP2171 ``Dynamic Wetting of Flexible, Adaptive, and Switchable Substrates'', projects HA-4382/11-1 and SE-3019/1-1. The authors gratefully acknowledge the Gauss Centre for Supercomputing e.V. (\url{www.gauss-centre.eu}) for funding this project by providing computing time on the GCS Supercomputer JUWELS~\cite{JUWELS} at Jülich Supercomputing Centre (JSC).
\end{acknowledgments}

\section*{Data availability}

The data that support the findings of this study are openly available in Zenodo at http://doi.org/10.5281/zenodo.6535952, reference number 6535952.





\bibliographystyle{ieeetr}
\bibliography{biblio}

\end{document}